# Effects of Impurity Scattering on Orbital Hall Conductivity and Orbital Transport in Ru-based Alloys

Yumin Yang[1,2], Yanxu Chen[1,2], Wenqi Xu[1,2] and Dahai Wei[1,2*]

[1] *State Key Laboratory of Semiconductor Physics and Chip Technologies, Institute of Semiconductors, Chinese Academy of Sciences, Beijing 100083, China*

[2] *College of Materials Science and Opto-Electronic Technology, University of Chinese Academy of Sciences, Beijing 100049, China*

E-mail: dhwei@semi.ac.cn

## ABSTRACT

The role of impurity scattering in the generation and transport of orbital current remains less established than in conventional spin Hall systems. Here we investigate Ru-based nonmagnet/ferromagnet bilayers in which the impurity scattering is tuned by Cu or Ti alloying. According to SOT measurement and thickness-dependent drift-diffusion analysis, we extract the effective orbital Hall conductivity and the orbital diffusion length. We find that the orbital Hall effect in polycrystalline Ru is dominated by intrinsic mechanism that is moderately robust against weak disorder but suppressed by stronger alloy disorder. However, the orbital diffusion length remains nearly unchanged at approximately 14 nm over the investigated impurity range. This behavior indicates that orbital transport is not governed simply by an impurity scattering. Together with previous temperature-dependent measurements, our results show that static impurities and dynamic lattice disorder affect orbital transport through distinct microscopic channels. These results provide new insight into how disorder governs orbital generation and transport, and offer experimental guidance for developing high-efficient orbitronic materials.

## I. INTRODUCTION

Angular momentum in solids is not carried only by spin. Bloch electrons also possess orbital angular momentum (OAM), whose nonequilibrium response is shaped by crystal symmetry, orbital hybridization, and coupling to the lattice. The resulting orbital currents have become the central object of orbitronics [1,2]. In particular, the orbital Hall effect (OHE) can generate transverse OAM currents from a charge current [3–6]. In addition, orbital injection into a ferromagnetic layer can generate orbital torques, providing a route toward efficient spin-orbit torque (SOT) switching [7–10]. Despite rapid progress in orbitronics, the mechanisms governing orbital-current generation, transport, and relaxation in nonmagnetic metals remain unresolved.

In widely used polycrystalline metallic films, the response of OAM to defects and disorder is still unclear. Experimental and theoretical studies have reported apparently conflicting pictures of OAM generation, transport, and relaxation. In an ideal crystal, the intrinsic OHE is commonly associated with orbital texture [5,6,11]. However, point defects, grain boundaries, and alloy disorder in sputtered polycrystalline films destroy crystal order and provide scattering centers, which may suppress the intrinsic OHE and generate extrinsic OHE contributions [12–14]. Nevertheless, the observation of sizable orbital torques in many polycrystalline films [15–20] suggests that the intrinsic OHE may be partly robust against disorder [21]. On the other hand, the transport length of OAM is equally unsettled. SOT measurements have reported OAM propagation over tens of nanometers [16,19], which may be explained by hotspots [22]; and THz ultrafast experiments have suggested long-range ballistic orbital transport [23,24]. By contrast, some THz-emission experiments [25,26] and some first-principles scattering calculations [27] have indicated much shorter orbital transport or relaxation lengths. Moreover, the OAM relaxation in metals has also been discussed in terms of Elliott-Yafet-like scattering [14], whereas Dyakonov-Perel-like orbital relaxation has also been reported [28,29]. These conflicting observations show that the influence of defects and scattering on OAM generation, transport, and relaxation remains poorly understood.

A controlled impurity-scaling experiment offers a way to address this problem. In spin Hall systems, resistivity scaling has been widely used to distinguish intrinsic, skew scattering, and side jump mechanisms, and to relate spin diffusion lengths to momentum scattering [30,31]. The analogous scaling rules for the OHE are not established. It remains unknown whether impurity scattering can enhance an extrinsic orbital Hall response, suppress an intrinsic orbital texture, shorten the orbital diffusion length, or leave OAM transport comparatively insensitive to atomic-scale disorder. It should be emphasized that spin transport models are often used as a starting point for describing orbital transport, but OAM and spin angular momentum have distinct microscopic natures [32,33]. Impurity-scattering experiments can therefore also clarify how spin and orbital transport differ.

In this work, we investigate these questions using Ru-based nonmagnet/ferromagnet (NM/FM) bilayers in which static impurity scattering is tuned by Cu or Ti alloying. Ru provides a useful orbital-current source because its spin Hall effect (SHE) is weak [34], while Cu and Ti introduce impurity scattering without adding a sizable spin Hall contribution [35–37]. By combining SOT measurement and thickness-dependent drift-diffusion analysis, we extract the effective orbital Hall conductivity and orbital diffusion length across Ru-Cu and Ru-Ti alloy series. We find that impurity scattering suppresses the effective orbital Hall conductivity but leaves the orbital diffusion length nearly unchanged. This separation between orbital-current generation and transport provides an experimental constraint on microscopic theories of OAM relaxation in disordered metals.

## II. EXPERIMENTAL METHODS

Ru-based NM/FM bilayers were prepared by magnetron sputtering at room temperature. Pure Ru and Ru alloys doped with Cu or Ti were deposited by co-sputtering. Ru was selected as the host material because it has a pronounced orbital Hall response but a negligible spin Hall response. Cu and Ti were selected as dilute-to-moderate alloying elements because their spin Hall effects are also expected to be weak. The alloy composition was tuned by adjusting the sputtering power applied to the Ru target and to the Cu or Ti target. In this study, we prepared pure Ru, a $Ru_{1-x}Cu_x$ series with nominal Cu concentrations x = 0.01, 0.03, 0.05, 0.10, and 0.15; and a $Ru_{1-x}Ti_x$ series with nominal Ti concentrations x = 0.05, 0.10, and 0.15. In addition, the Ru films sputtered at room temperature already contain defects and grain boundary scattering; thus 0.1at.% impurity level is not necessary. All the films were deposited on sapphire substrates with a $SiO_2$(10)/Ti(2) buffer and Ti(1)/$SiO_2$(10) capping, where the numbers in parentheses denote thicknesses in nanometers and the order follows the deposition sequence. The amorphous $SiO_2$ layers were used to break the surface structure of the sapphire, and the thin Ti layers were used to improve adhesion. For each alloy composition, the NM-thickness series was prepared in a single deposition run using a movable shutter, ensuring that samples with different NM thicknesses shared identical growth conditions. After film growth, devices were fabricated using standard photolithography, ion beam etching, and lift-off processes.

## III. RESULTS AND DISCUSSION

### A. Basic transport properties of pure and alloyed Ru heterostructures

We first characterize the electrical transport properties of the pure Ru and Ru-alloy layers. Fig. 1a shows the device conductance as a function of NM thickness $t_{\mathrm{NM}}$ for pure Ru and for the alloyed series. All series exhibit approximately linear conductance-thickness behavior over the thickness range studied. This indicates that the NM resistivity is nearly thickness independent, excluding a sizable current density gradient or local current-vorticity contribution in the NM layer. The different slopes of the linear fits therefore reflect the different resistivities of pure Ru and of the Cu- or Ti-alloyed Ru layers. At room temperature, the resistivity of pure Ru is approximately 22 μΩ cm. Cu alloying increases the resistivity from approximately 25.9 μΩ cm for $Ru_{0.99}Cu_{0.01}$ to 71.3 μΩ cm for the highest-Cu alloy. Ti alloying also increases the resistivity, from 45.6 μΩ cm for $Ru_{0.95}Ti_{0.05}$ to 75.2 μΩ cm for $Ru_{0.85}Ti_{0.15}$. These results show that both Cu and Ti introduce substantial additional scattering into the Ru host.

To separate the temperature-dependent and residual contributions to the resistivity, we measured the resistivity from 10 K to 300 K using a four-probe method. Fig. 1b shows the $\rho(T)$ curves for the $Ru_{1-x}Cu_x$ alloys. For all Cu concentrations, $\rho$ decreases upon cooling and becomes nearly temperature independent below approximately 30 K. Following Matthiessen's rule analysis, the low temperature residual resistivity is assigned mainly to static scattering from impurities, defects, and grain boundaries, whereas the reduction between 300 K and the residual value is taken as the phonon-related contribution.

Fig. 1c and d summarize the phonon-related contribution $\rho_{\text{phon}}$ and the residual contribution $\rho_{\text{imp}}$ for pure Ru and for the alloyed Ru samples. Compared with pure Ru, alloying reduces $\rho_{\text{phon}}$. This reduction should be understood as a decrease in the temperature-dependent part of the resistivity, rather than as direct evidence for weaker phonon scattering. It may reflect the suppression of phonon induced resistivity by strong impurity scattering, or a breakdown of a simple additive Matthiessen relation in the strongly disordered regime. By contrast, $\rho_{\text{imp}}$ changes much more strongly with dopant concentration, confirming that alloying efficiently enhances static scattering. The Cu series varies less smoothly with nominal composition than the Ti series, which is likely related to uncertainty in the actual Cu concentration because the Cu deposition rate is difficult to control accurately at low sputtering power.

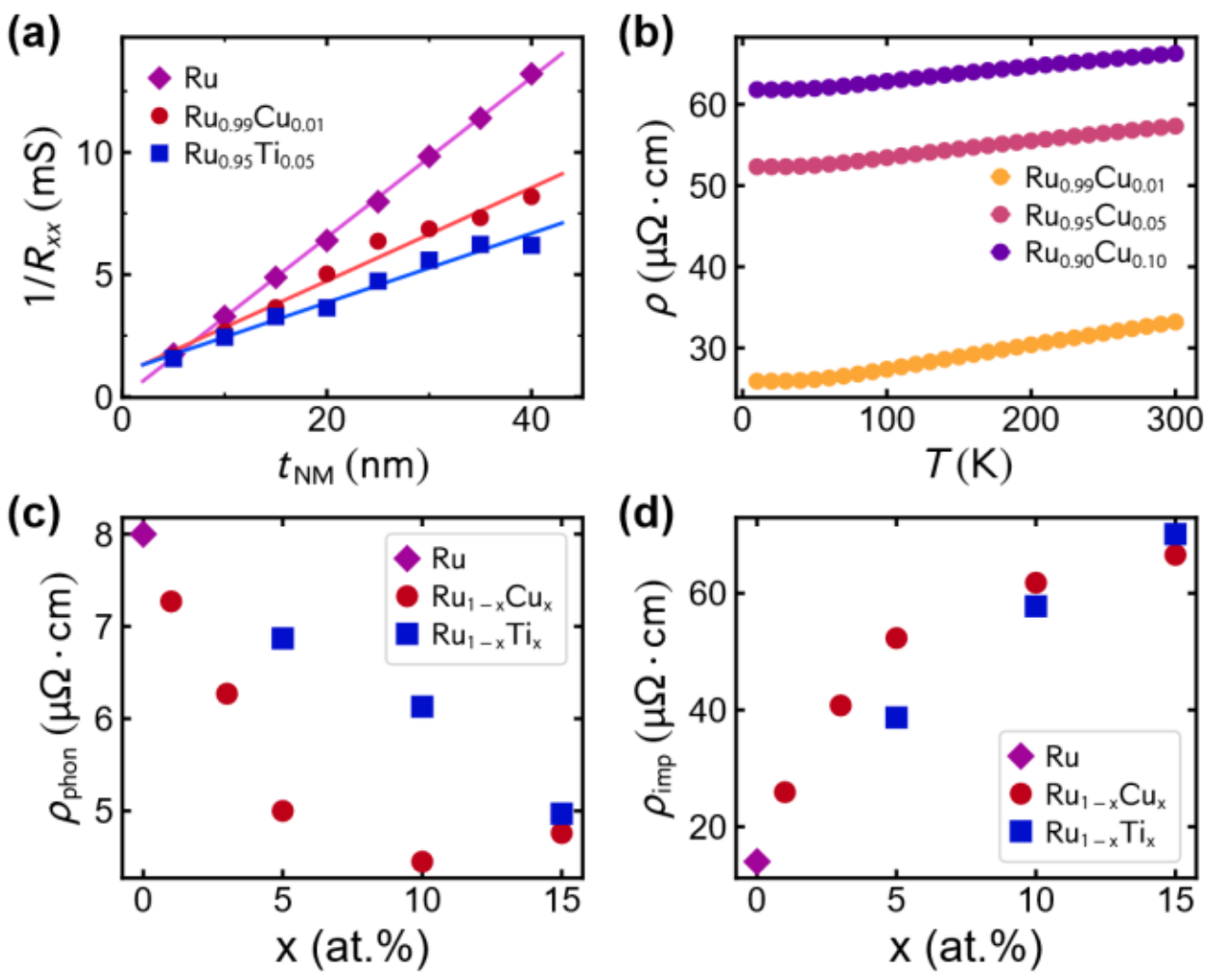

Fig. 1. Basic transport properties of pure and alloyed Ru heterostructures. (a) Device conductance $1/R_{xx}$ as a function of NM thickness $t_{\text{NM}}$. (b) Resistivity $\rho$ of $Ru_{1-x}Cu_x$ alloys as a function of temperature $T$. (c) Phonon-related contribution and (d) impurity-related contribution to the resistivity in pure Ru and Ru alloys.

### B. ST-FMR measurement of orbital torque in Ru-alloy/Co bilayers

We used spin-torque ferromagnetic resonance (ST-FMR) to measure the orbital torque and thereby evaluate the strength of the OHE. As illustrated in Fig. 2a, a radio frequency (RF) current is applied along the microstrip, while an in-plane magnetic field $H$ is applied at an oblique angle $\varphi$ with respect to the current direction. The resulting rectification voltage $V_{\text{DC}}$ is detected using a lock-in amplifier. Fig. 2b shows the field-dependent $V_{\text{DC}}$ measured at different RF frequencies $f$. The resonance field $H_{\text{res}}$ shifts to higher field as $f$ increases. Fig. 2c shows a representative ST-FMR signal for a $Ru_{90}Cu_{10}$(20)/Co(5) device, together with the Lorentzian fit. A pronounced symmetric component $V_S$ is observed, indicating a sizable damping-like effective field $H_{DL}$, whereas the antisymmetric component $V_A$ mainly arises from the field-like effective field and Oersted field $H_{FL,Oe}$. For quantitative evaluation of the orbital torque, the conventional $V_S/V_A$ ratio analysis is not sufficiently reliable because the orbital torque depends on FM thickness $t_{FM}$. We therefore directly extract the $H_{DL,y}$ component of the SOT effective field from the ST-FMR signal [38,39]:

$$V_S = \frac{I_{RF}\Delta R \sin 2\varphi \cos\varphi}{2\alpha(2H_{res}+M_{eff})} \frac{\Delta_H^2}{(H-H_{res})^2+\Delta_H^2} H_{DL}$$

$$V_A = \frac{I_{RF}\Delta R \sin 2\varphi \cos\varphi}{2\alpha(2H_{res}+M_{eff})} \frac{\Delta_H(H-H_{res})}{(H-H_{res})^2+\Delta_H^2} \sqrt{1+\frac{M_{eff}}{H_{res}}}\, H_{FL,Oe} \quad (1)$$

where $\Delta_H$ is the half-width-at-half-maximum linewidth of the Lorentzian line shape. The effective magnetization $M_{eff}$ is obtained from the $f$-dependence of the resonance field by fitting the Kittel formula $2\pi f = \gamma\sqrt{H_{res}(H_{res}+M_{eff})}$, and $\gamma$ is the gyromagnetic ratio. And the Gilbert damping coefficient $\alpha$ can be fitted by $\Delta_H = 2\pi f\alpha/\gamma + \Delta_{H,0}$. The in-plane angular-dependent magnetoresistance amplitude $\Delta R$ is extracted by fitting $R(\varphi) = R_0 + \Delta R\cos^2\varphi$. The RF current amplitude $I_{RF}$ is calibrated from Joule heating by comparing the resistance change $\Delta R_{heat}$ induced by a DC current $I_{DC}$ with that induced by the RF current. We further analyzed the $\varphi$ dependence of $V_S$. As shown in Fig. 2d, the signal is dominated by the $\sin 2\varphi\cos\varphi$ expected for $H_{DL,y}$, while the $\sin 2\varphi$ term attributed mainly to the net Oersted field generated by the asymmetric current distribution is small.

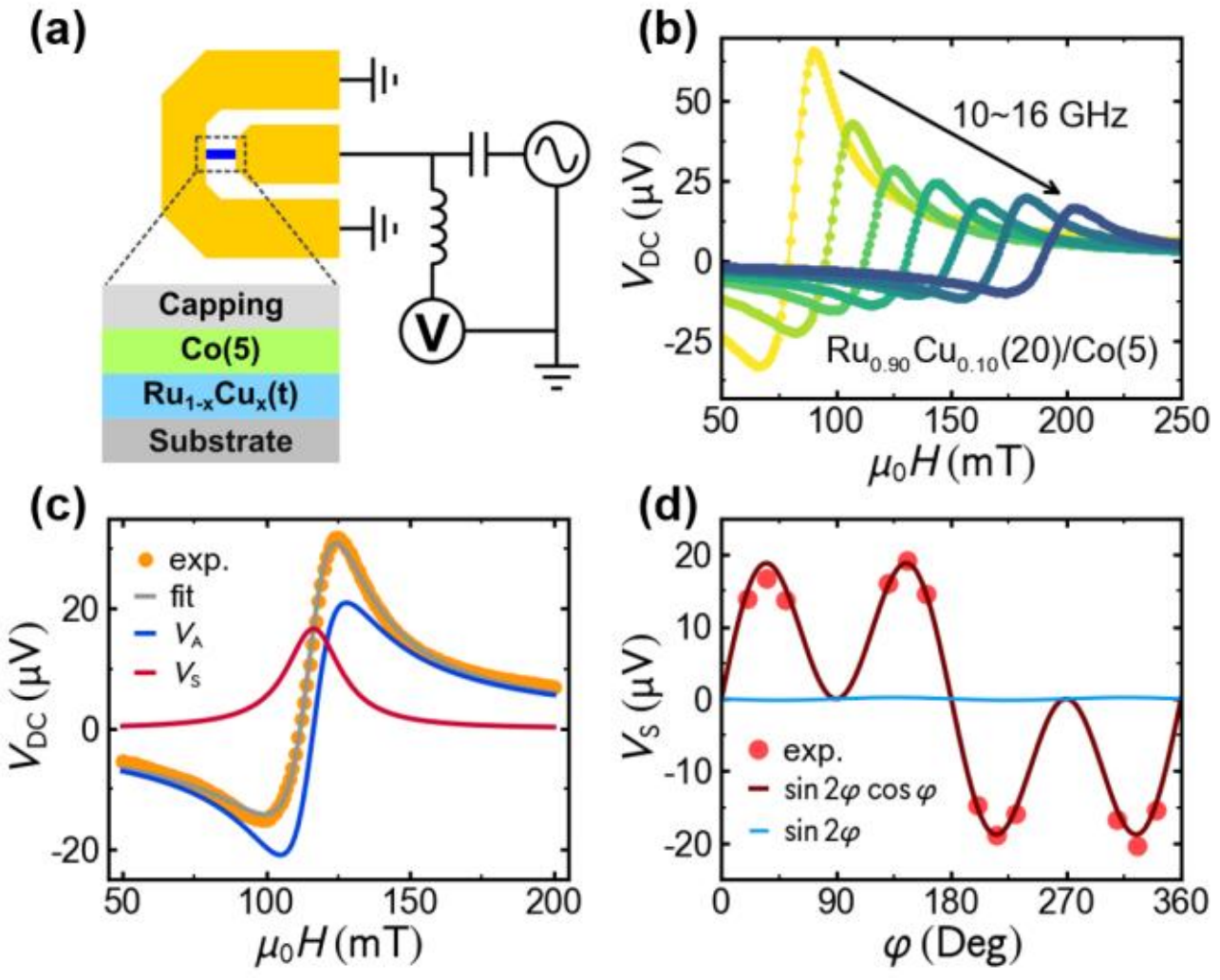

Fig. 2. ST-FMR measurement. (a) Schematic of the ST-FMR measurement geometry. (b) Field-dependent rectification voltage $V_{DC}$ measured at different RF frequencies, (c) decomposition of $V_{DC}$ into symmetric $V_S$ and antisymmetric $V_A$ Lorentzian components at 12 GHz and $\varphi$ = 35 deg, and (d) angular dependence of $V_S$ for $Ru_{0.90}Cu_{0.10}$(20)/Co(5) sample.

### C. Effective orbital Hall conductivity and orbital diffusion length

To quantify the orbital torque, we use the electric-field-normalized damping-like SOT efficiency,

$$\xi_{\mathrm{DL}}^{\mathrm{E}} = \frac{2e}{\hbar}\frac{\mu_0 M_{\mathrm{s}} t_{\mathrm{FM}} H_{\mathrm{DL}}}{E} \tag{2}$$

where $M_s$ is the saturation magnetization, $t_{\mathrm{FM}}$ is the FM thickness, $H_{\mathrm{DL}}$ is the damping-like effective field, and $E$ is the electric field in the device, estimated from $I_{\mathrm{RF}}$, $R_0$, and the device length $l$. Because the measured torque depends on both orbital current generation in the NM layer and orbital diffusion toward the Co interface, measurements on a single NM thickness cannot separate these two contributions. We therefore fit the $t_{\mathrm{NM}}$ dependence of $\xi_{\mathrm{DL}}^{E}$ using a drift-diffusion model to extract the effective orbital Hall conductivity $\sigma_{\mathrm{OH}}^{\mathrm{eff}}$ and the orbital diffusion length $\lambda_{\mathrm{NM}}$:

$$\xi_{\mathrm{DL}}^{\mathrm{E}} = \sigma_{\mathrm{OH}}^{\mathrm{eff}}\left(1 - \mathrm{sech}\frac{t_{\mathrm{NM}}}{\lambda_{\mathrm{NM}}}\right) + \xi_{\mathrm{DL,0}}^{E} \tag{3}$$

where $\xi_{\mathrm{DL,0}}^{E}$ accounts for the self-torque and interfacial contribution in the zero-thickness limit. Fig. 3a shows the $t_{\mathrm{NM}}$-dependent $\xi_{\mathrm{DL}}^{E}$ for $Ru_{1-x}Cu_x$ alloys, while Fig. 3b shows the corresponding data for $Ru_{1-x}Ti_x$ alloys, with $x$ = 0~0.15 including pure Ru. In both alloy series, $\xi_{\mathrm{DL}}^{E}$ increases with $t_{\mathrm{NM}}$ and gradually approaches saturation at larger thickness. This common thickness dependence indicates that the damping-like torque is dominated by orbital current generated in the Ru-based NM layer and transported to Co, rather than by a purely interfacial torque.

The drift-diffusion fits reproduce the data for both Cu- and Ti-alloyed series. Although the absolute magnitude of $\xi_{\mathrm{DL}}^{E}$ decreases as the alloy concentration and residual resistivity increase, the overall thickness-dependent shape remains nearly unchanged. For the Cu series, the fitted $\lambda_{\mathrm{NM}}$ remains approximately 12~16 nm across $x$ = 0~0.15, while $\sigma_{\mathrm{OH}}^{\mathrm{eff}}$ decreases markedly once the Cu concentration reaches about 5 at.%. A similar behavior is observed for the Ti series: $\sigma_{\mathrm{OH}}^{\mathrm{eff}}$ decreases with increasing Ti concentration, whereas $\lambda_{\mathrm{NM}}$ remains close to 14 nm. Thus, Cu and Ti impurity scattering suppress the orbital-current generation efficiency in Ru but do not appreciably shorten the OAM transport length.

This robustness is further supported by the normalized thickness dependence in Fig. 3c. Data from pure Ru, $Ru_{1-x}Cu_x$, and $Ru_{1-x}Ti_x$ collapse onto a common trend that is described by a single diffusion length of about 14.4 nm. The collapse shows that the characteristic diffusion profile is nearly insensitive to impurity species and concentration. Therefore, the orbital diffusion length in Ru-based alloys is not governed simply by the impurity-limited momentum mean free path.

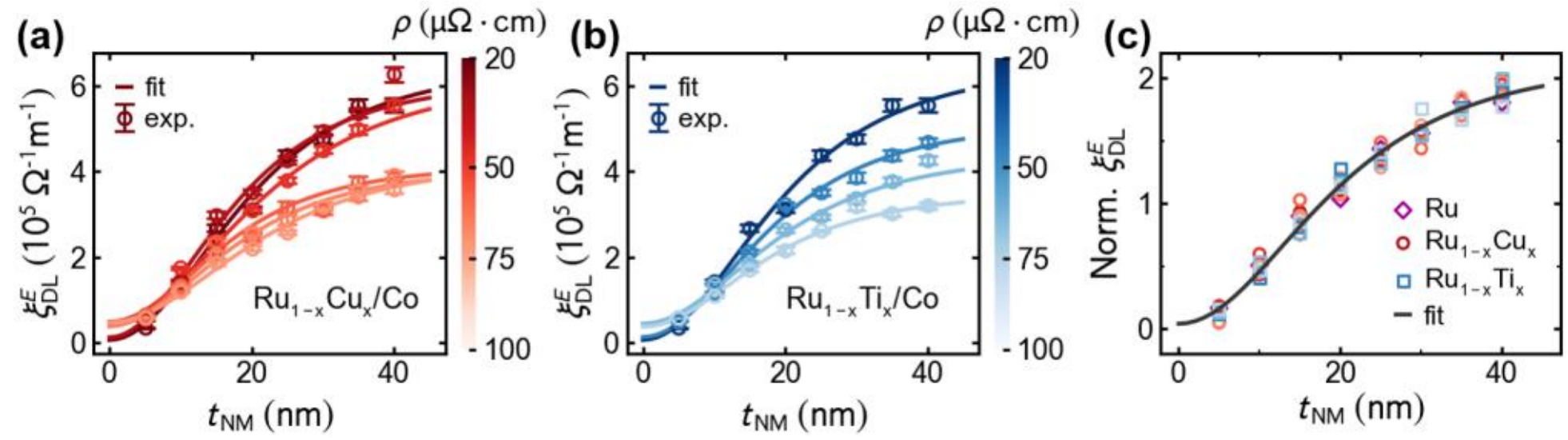

Fig. 3. Thickness dependence of the electric-field-normalized damping-like SOT efficiency. $\xi_{\mathrm{DL}}^{E}$ as a function of NM thickness $t_{\mathrm{NM}}$ for (a) $Ru_{1-x}Cu_x$ and (b) $Ru_{1-x}Ti_x$ alloys with different resistivity, where $x = 0\sim0.15$. (c) Normalized $\xi_{\mathrm{DL}}^{E}$ as a function of $t_{\mathrm{NM}}$ for all series, together with a fit.

## D. Scaling analysis with impurity resistivity

The thickness-dependent analysis shows that impurity scattering affects orbital generation and orbital transport differently. We now examine this distinction more explicitly by scaling the extracted parameters with impurity resistivity. Fig. 4a plots $\lambda_{\mathrm{NM}}$ as a function of the residual impurity contribution $\rho_{\mathrm{imp}}$ for pure Ru, $Ru_{1-x}Cu_x$, and $Ru_{1-x}Ti_x$. Although Cu and Ti alloying substantially increases $\rho_{\mathrm{imp}}$, $\lambda_{\mathrm{NM}}$ remains within approximately 12~16 nm and shows no monotonic dependence on $\rho_{\mathrm{imp}}$. Thus, the OAM transport length is nearly insensitive to static impurity scattering over the present disorder range.

Because the full three-parameter fits produce larger scatter owing to experimental uncertainty, we also refit the thickness-dependent data with $\lambda_{\mathrm{NM}}$ fixed at 14.4 nm, as suggested by the normalized diffusion profile in Fig. 3c. In this procedure, only $\sigma_{\mathrm{OH}}^{\mathrm{eff}}$ and $\xi_{\mathrm{DL},0}^{E}$ are varied to fit the $t_{\mathrm{NM}}$ dependence of $\xi_{\mathrm{DL}}^{E}$. The resulting $\sigma_{\mathrm{OH}}^{\mathrm{eff}}$ is shown in Fig. 4b; the directly fitted values are provided in the Supplemental Material and show the same trend. In the high conductivity regime, represented by pure Ru and lightly alloyed Ru, $\sigma_{\mathrm{OH}}^{\mathrm{eff}}$ changes only weakly with $\sigma_{\mathrm{NM}}$. When $\sigma_{\mathrm{NM}}$ falls below approximately $2.5 \times 10^{6}\ \Omega^{-1}\ \mathrm{m}^{-1}$, $\sigma_{\mathrm{OH}}^{\mathrm{eff}}$ decreases markedly, indicating that stronger alloy disorder suppresses orbital generation. This low conductivity regime should not be identified simply as skew scattering dominated, because impurity scattering can also disrupt the intrinsic orbital texture and orbital hybridization that support the OHE. Fig. 4c shows that the effective orbital Hall angle $\theta_{\mathrm{OH}}$ increases from about 0.15 to 0.25 with increasing longitudinal resistivity $\rho_{xx}$ and then tends to saturate. This trend resembles spin Hall systems and indicates that a moderate increase in resistivity can improve the SOT efficiency even when $\sigma_{\mathrm{OH}}^{\mathrm{eff}}$ is reduced.

This two-regime scaling behavior has some similarity to conventional spin Hall scaling, but the difference between Ru OHE and Pt SHE is important. In many Pt-based alloys, a strong reduction of the intrinsic spin Hall response usually requires heavier alloying, such as dopant concentrations of order 20 at.% or strong composite disorder [30]. In the present $Ru_{1-x}Cu_x$ and $Ru_{1-x}Ti_x$ alloys, however, $\sigma_{\mathrm{OH}}^{\mathrm{eff}}$ already decreases substantially when the impurity concentration reaches about 5 at.%. This suggests that the intrinsic OHE in Ru is more sensitive to impurity scattering or local atomic disorder. The difference may originate from the distinct microscopic origins of the two effects: the intrinsic SHE in Pt is mainly associated with spin Berry curvature generated by strong SOC, whereas the OHE in Ru depends directly on orbital texture and orbital hybridization, which can be disturbed by relatively dilute impurities.

The weak impurity dependence of $\lambda_{\mathrm{NM}}$ also contrasts with standard spin relaxation expectations. In an Elliott-Yafet-like picture, stronger momentum scattering would shorten the angular-momentum lifetime and reduce the diffusion length. A Dyakonov-Perel-like picture would instead predict an increase in lifetime under stronger scattering. The present data show neither trend. The Ru-alloy data therefore separate two effects that are often coupled in spin Hall systems: impurity scattering reduces $\sigma_{\mathrm{OH}}^{\mathrm{eff}}$ but leaves $\lambda_{\mathrm{NM}}$ nearly unchanged. A possible physical picture is sketched in Fig. 4d. The current induced OAM in Ru may be carried by itinerant electronic states with an extended wave-packet nature, rather than by localized atomic orbitals. If the relevant orbital wave packet extends over many lattice sites, a local atomic-scale impurity will not destroy the coherence of the extended OAM. This picture naturally explains why the residual resistivity is strongly modified by Cu or Ti alloying, whereas the orbital diffusion length remains robust. In addition, it also leaves room for electron-lattice coupling model to explain orbital transport, in which OAM relaxation is not controlled by a single-particle impurity scattering time alone.

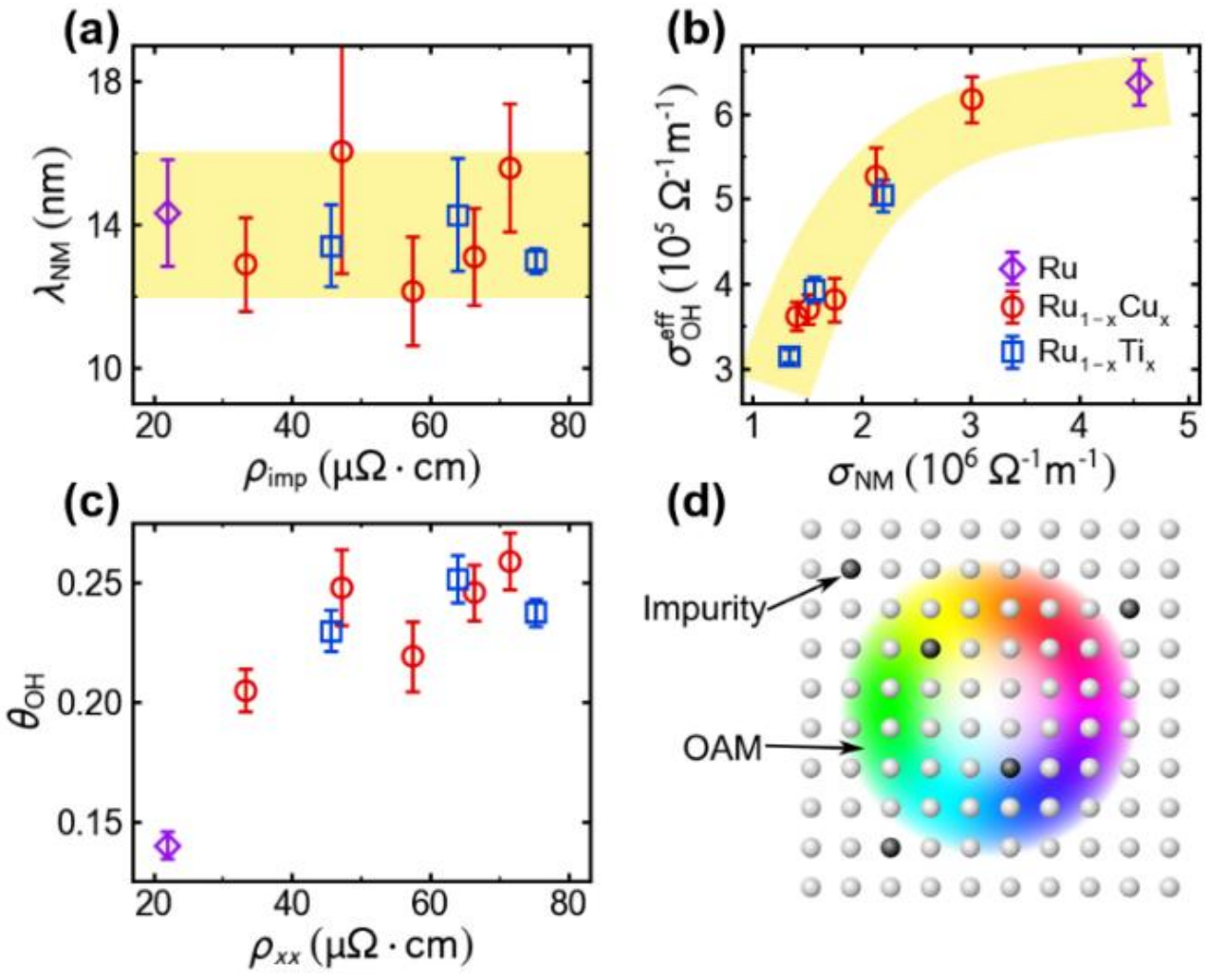

Fig. 4. Scaling analysis with impurity scattering. (a) Orbital diffusion length $\lambda_{\mathrm{NM}}$ as a function of impurity-related residual resistivity $\rho_{\mathrm{imp}}$. (b) Effective orbital Hall conductivity $\sigma_{\mathrm{OH}}^{\mathrm{eff}}$ as a function of NM conductivity $\sigma_{\mathrm{NM}}$ obtained with fixed $\lambda_{\mathrm{NM}}$=14.4 nm. (c) Effective orbital Hall angle $\theta_{\mathrm{OH}}$ as a function of longitudinal resistivity $\rho_{xx}$. (d) Schematic illustration of orbital-current transport, where the color wheel represents the phase of orbital angular momentum.

Moreover, this impurity-scaling behavior is qualitatively different from our previous temperature-dependent Ru/Co measurements [40]. In that case, reducing phonon scattering at low temperature enhanced both the damping-like efficiency and the fitted orbital diffusion length, with $\lambda_{\mathrm{NM}}$ increasing from about 15 nm to about 20 nm and $\sigma_{\mathrm{OH}}^{\mathrm{eff}}$ also increasing. By contrast, static impurity scattering in the present alloys strongly changes the residual resistivity but does not appreciably shorten $\lambda_{\mathrm{NM}}$. The comparison suggests that dynamic lattice disorder and static impurity disorder affect orbital transport differently. Phonons may affect more directly to orbital coherence or to angular-momentum transfer across the Ru/Co interface [8,41–43], whereas static impurities mainly perturb orbital-current generation in the Ru layer. Therefore, the temperature dependence of $\sigma_{\mathrm{OH}}^{\mathrm{eff}}$ should not be interpreted solely as evidence for a conventional skew-scattering contribution.

We also note the recent study of Mn thin films [14]. That work found that the OHE and orbital relaxation can be quantified from strongly disordered Mn films to single-crystalline α-Mn. In the high resistivity, hopping dominated regime, the $\sigma_{\mathrm{OH}}^{\mathrm{eff}}$ scales approximately linearly with the $\sigma_{\mathrm{NM}}$, while the orbital relaxation time remains on the order of picoseconds and varies with the degree of structural order. However, the two observations refer to different transport regimes and therefore are not contradictory. The Mn data in the strongly disordered regime would correspond to the low $\sigma_{\mathrm{NM}}$ and low $\sigma_{\mathrm{OH}}^{\mathrm{eff}}$ corner of the scaling plot in Fig. 4b, whereas our Ru alloys remain in a comparatively high $\sigma_{\mathrm{NM}}$ metallic regime. In hopping dominated Mn, strong scattering can suppress the intrinsic orbital response, which may also explain why the experimentally inferred $\sigma_{\mathrm{OH}}^{\mathrm{eff}}$ is much smaller than the theoretical value. The comparison therefore reinforces, rather than contradicts, the conclusion that the impurity dependence of the OHE is strongly regime dependent.

## IV. CONCLUSION

We have investigated how impurity scattering affects orbital-current generation and transport in Ru-based NM/FM heterostructures. By alloying Ru with Cu or Ti, we systematically increase the residual resistivity and establish a controlled static-scattering axis for scaling analysis. The $t_{\mathrm{NM}}$-dependence of the $\xi_{\mathrm{DL}}^{E}$ is well described by a drift-diffusion model, allowing us to extract $\sigma_{\mathrm{OH}}^{\mathrm{eff}}$ and $\lambda_{\mathrm{NM}}$. With increasing impurity scattering, the $\sigma_{\mathrm{OH}}^{\mathrm{eff}}$ first changes only weakly and then decreases rapidly once the alloy concentration exceeds about 5 at.%, whereas the $\lambda_{\mathrm{NM}}$ remains close to 14 nm across both Cu- and Ti-alloyed series. These results indicate that the OHE in polycrystalline Ru is dominated by an intrinsic OHE mechanism that is moderately robust against weak structural disorder but is suppressed when disorder becomes sufficiently strong. Moreover, compared with the SHE in typical Pt-based system, the OHE in Ru has stronger disorder sensitivity. At the same time, the nearly impurity-independent orbital diffusion length suggests that OAM transport in the present high conductivity metallic regime is not governed simply by an impurity-limited momentum mean free path. A possible explanation is that the current-induced OAM is carried by itinerant electrons with a spatially extended wave packet, making orbital transport comparatively tolerant of local atomic-scale impurities.